\documentclass[aps,pra,showpacs,superscriptaddress,twocolumn]{revtex4}
\usepackage{epsfig}
\usepackage{graphicx}
\usepackage{amsmath,amssymb}
\usepackage{epstopdf}

\begin{document}
\title{Weakly bound states of two- and three-boson systems in the crossover from two to three dimensions}
\author{M.~T. Yamashita}
\affiliation{Instituto de F\'\i sica Te\'orica, UNESP - Univ Estadual Paulista, C.P. 70532-2, CEP 01156-970, S\~ao Paulo, SP, Brazil}
\author{F.~F. Bellotti}
\affiliation{Instituto Tecnol\'{o}gico de Aeron\'autica, 12228-900, S\~ao Jos\'e dos Campos, SP, Brazil}
\affiliation{Department of Physics and Astronomy, Aarhus University, DK-8000 Aarhus C, Denmark}
\affiliation{Instituto de Fomento e Coordena\c{c}{\~a}o Industrial, 12228-901, S{\~a}o Jos{\'e} dos Campos, SP, Brazil}
\author{T. Frederico}
\affiliation{Instituto Tecnol\'{o}gico de Aeron\'autica, 12228-900, S\~ao Jos\'e dos Campos, SP, Brazil}
\author{D.~V. Fedorov}
\author{A.~S. Jensen}
\author{N.~T. Zinner}
\affiliation{Department of Physics and Astronomy, Aarhus University, DK-8000 Aarhus C, Denmark}
\date{\today }
\begin{abstract}
The spectrum and properties of quantum bound states is strongly 
dependent on the dimensionality of space. How this comes about and how 
one may theoretically and experimentally study the interpolation between 
different dimensions is a topic of great interest in different fields of 
physics. 
In this paper we study weakly bound states of non-relativistic two and three boson 
systems when passing continuously from a three (3D) to a two-dimensional (2D) 
regime within a 'squeezed dimension' model.  We use periodic boundary conditions 
to derive a surprisingly simple form of the three-boson Schr{\"o}dinger
equation in momentum space that we solve numerically.
Our results show a distinct dimensional crossover as three-boson states 
will either disappear into the continuum or merge with a 2D counterpart, 
and also a series of sharp transitions in the ratios of three-body and two-body 
energies from being purely 2D to purely 3D.
\end{abstract}
%03.65.Ge Solutions of wave equations: bound states
%36.10.-k Exotic atoms and molecules
%21.45.-v Few-body systems
\maketitle

\section{Introduction}
Strongly interacting quantum particles are of great importance in 
many fields including nuclear and particle physics, 
condensed-matter physics, and atomic and molecular physics. 
However, they can be very hard to approach due to their
intrinsically non-perturbative nature and often it is 
necessary to employ numerical simulations such as Monte
Carlo \cite{drut2013} and/or lattice field theory 
techniques as done with great success in lattice QCD \cite{latqcd}. 
In the limit of few particles, strongly interacting systems 
can display truly remarkable features such as the Efimov
effect \cite{ef70} where a geometric series of three-body bound
states of three bosons occur at the threshold for the binding
of any two-body subsystem. This is a key insight into the 
often counterintuitive behavior of few-body systems that is 
extremely difficult to capture without analytical guidance. 
Moreover, it is an effect that is intimately tied to the 
dimensionality of space as it will not happen in two
dimensions.

Cold atomic gases have proven their ability as excellent
quantum simulation tools due to the tunability of interactions, 
geometry, and inter-particle statistical properties \cite{coldatoms1,coldatoms2,zinjen2013}.
A recent frontier is the study of strongly interacting atomic Bose gases
in three \cite{papp2008,pollack2009,navon2011,wild2012,fletcher2013,rem2013,makotyn2014} and
most recently in two dimensions \cite{na2013,makhalov2014}.
At the few-body level, three-body states linked to the
Efimov effect have been observed in three 
dimensions \cite{kraemer2006,ottenstein2008,pollack2009b,zaccanti2009,gross2009,huckans2009,williams2009,lompe2010a,gross2010,lompe2010b,nakajima2010,
nakajima2011,berninger2011,machtey2012a,machtey2012b,knoop2012,hulet2013,roy2013,tung2014,huang2014,pires2014}
using a variety of different atomic species and two-body 
Feshbach resonances \cite{chin2010}. In spite of the tunability 
of the external trapping geometry of cold atomic systems, there has
been little study of how the three-boson
bound state problem undergoes its dramatic change from displaying the 
Efimov effect in three dimensions to having only two bound states in two dimensions
\cite{tjon}. A key question is 
whether it is possible to interpolate these limits in simple theoretical 
terms and subsequently explore this in simulations using 
both more involved numerical methods and experimental setups.

Here we present a model that has the ability to interpolate
geometrically between two and three spatial dimensions and thus study
this important crossover for both two- and three-body bound states 
of identical bosons. A 'squeezed' dimension is employed with periodic 
boundary conditions (PBC) whose size can be varied to interpolate the two 
limits. It has the unique feature that it can be 
regularized analytically which is a great advantage for its numerical 
implementation allowing us to go smoothly between both limits. While 
PBC is not typically used in few-body calculations, it is nevertheless
a standard trick when addressing larger systems \cite{drut2013,endres2013,rossi2014}
and our results may thus also serve as a benchmark. 
Since the seminal work of Wilson \cite{wilson1975} and t'Hooft \cite{thooft1973}, 
using the dimensionality of a given model as a parameter has been 
a standard tool in high-energy and condensed matter physics. More recently
such techniques have been applied to strongly interacting Fermi gases \cite{nishida2010}
in the context of cold atoms. Moreover, in the realm of few-body physics  
mixed dimensional systems are promising for extension of the Efimov scenario
to new setups \cite{nishida2011}.

For many experimental setups in cold atoms, the transverse confining geometry
is given by a harmonic trapping potential and a very interesting recent theoretical 
study \cite{levinsen2014} has considered the properties of three-boson states 
under strong transverse confinement. The very recent 
successful production of box potential traps for bosons
\cite{gaunt2013,schmidutz2014} means that open (hard wall) boundary
conditions (OBC) are now also accessible. This still leaves the question of how to 
realize periodic boundary conditions in a cold atomic gas experiment. 
However, we would not expect large qualitative differences between 
OBC and PBC and more likely only quantitative changes.
This is of course also one of the reasons for employing periodic boundary
conditions in most contexts throughout physics. The 
theoretical elegancy and tractability of calculations in the three-body
system is the strong incentive that we have for pursuing this geometry
even if at present an experimental realization has not been found.

\section{Dimer energy with periodic boundary condition} 
In our model we will assume
periodic boundary conditions along one direction (chosen to be the $z$-axis) 
which is initially a 
condition on the single-particle coordinates. We will now
argue, however, that it may be implemented in the relative
momentum which significantly simplifies its implementation.
Let $p_1$ and $p_2$ be the $z$-components of the momenta of 
particles 1 and 2 with coordinates $z_1$ and
$z_2$, respectively. Then we may rewrite
\begin{align}
p_1 z_1+p_2 z_2=Zp_{CM}+zp_z,
\end{align}
where $p_{CM}=(p_1+p_2)/\sqrt{2}$ and $p_z=(p_1-p_2)/\sqrt{2}$, while $Z=(z_1+z_2)/\sqrt{2}$
and $z=(z_1-z_2)/\sqrt{2}$. PBC implies that we must quantize
according to $p_1=2\pi n_1/L$ and $p_2=2\pi n_2/L$, where $L$
is the length of the periodic dimension. We thus see that $p_{CM}$
and $p_z$ will in turn also be quantized as a sum or a difference
of a pair of integers. Furthermore, the Hamiltonian separates in 
CM and relative momenta. Thus we may treat $p_{CM}$ and $p_z$ 
independently. 

Disgarding the center-of-mass momentum we may now concentrate 
on the relative part. The relative momenta along the plane 
are then given by $\vec p_\perp=(p_x,p_y)$, while
\begin{eqnarray}
p_z=\frac{\sqrt{2}\pi n}{L}=\frac{n}{R} \ , \label{eq1}
\end{eqnarray}
with $n=(n_1-n_2)=0,\pm1,\pm2...$ and $L=\sqrt{2}\pi R$. 
The length of the squeezed dimension corresponds to a 
radius, $R$, that interpolates between the 2D limit for $R\to 0$ and  
the 3D limit for $R\to \infty$.
In this paper we will use a contact (zero-range) 
interaction to study the continuous 
transition from 3D to 2D regimes. 

\begin{figure}[!thb]
\centering
\includegraphics[scale=0.33,clip=true]{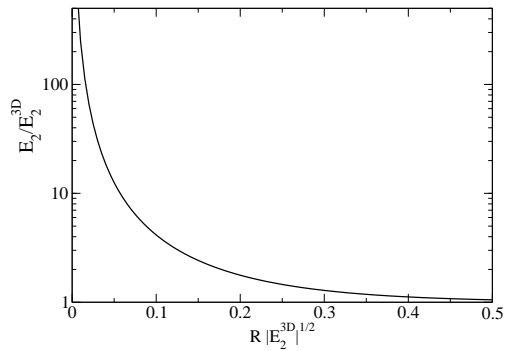}
\caption{Dimer energies $E_2/E_2^{3D}$ as a function of $R\sqrt{|E_2^{3D}|}$ for periodic boundary 
conditions.}\label{e2fig}
\end{figure}

Here we consider the case where
we allow $E_2$ to vary with $R$ under the physical condition that 
the magnetic field is fixed and thus we have a fixed dimer energy
in 3D,  $E_2^{3D}$. 
This implies that the two-body T-matrix in the limit 
$R\to\infty$ has to recover a pole exactly at $E_{2}^{3D}$.
This means that for finite $R$ we must solve
\begin{eqnarray}
\int {d^3p\over E_2^{3D}-
p^2} -
\frac{1}{R}\sum_n\int {d^2p_\perp\over E_2-
p_\perp^2-\frac{n^2}{R^2}}
=0 . \label{rtau1}
\end{eqnarray}
As both terms in Eq.~\eqref{rtau1} are divergent, ultraviolet cutoffs 
that are consistent with the correct 3D limit must be introduced. 
It is enough to regularize the transverse momentum 
integral $d^2p_\perp$ in both terms of (\ref{rtau1}) with a cutoff 
$\Lambda$ and then take the limit $\Lambda\to\infty$. 
\begin{multline}
\lim_{\Lambda\to\infty} \left\{\int^{\infty}_{-\infty} dy  
\ln \left[{-E_2^{3D}R^2+y^2 \over -E_2^{3D}R^2+y^2+(\Lambda\,R)^2 }\right] \right. \\ \left.
- \sum_{n=-\infty}^{\infty}  \ln \left[{E_2\,R^2-n^2\over E_2\,R^2-n^2-(\Lambda\,R)^2 }\right]\right\}
=0 \ , \label{rtau4}
\end{multline}
where $y\equiv R\,p$. Performing the analytical integration and the sum we have
\begin{multline}
\lim_{\Lambda\to\infty} \left\{ \pi\, R\left(\sqrt{-E_2^{3D}}- \sqrt{-E_2^{3D}+\Lambda^2}\right) \right. \\ \left. - \ln\left({\sinh\pi\sqrt{-E_2}R\over
\sinh\pi\sqrt{-E_2+\Lambda^2}R}\right)\right\} \\
= \pi\, R\sqrt{-E_2^{3D}}- \ln
\left(2\sinh\pi\sqrt{-E_2}R\right)=0\ . \label{rtau5}
\end{multline}
By recognizing that $\sqrt{-E_2^{3D}}\to1/a$ in the zero-range limit (where $a$ is the two-body scattering 
length), the energy of the dimer is
\begin{equation}
\sqrt{-E_2}
= \frac{1}{\pi R}\sinh^{-1} \frac{e ^{\pi\, R/a}}{2} \ ,
\label{rtau7}
\end{equation}
The energy of the dimer is shown in Fig.~\ref{e2fig}.
For $R\to 0$ it goes to $\sqrt{-E_2}\sim (\pi R)^{-1}\,  \sinh^{-1} \frac{1}{2}  = 0.153174\,  
R^{-1}$, which does not depend on the scattering length. Therefore, for any 3D two-body 
subsystem - bound or virtual - a strong deformation of the trap towards 2D always 
binds the dimer with an energy given by the trap energy scale ($1/R^2$ in this case). 
In the unitary limit where $a\to\infty$ we find this bound state energy for any finite
$R$. This is analogous to the famous quasi-2D harmonic trap result of Petrov and 
Shlyapnikov \cite{petrov2001}.
In general, the precise way in which a low-dimensional geometry
is obtained will be reflected in the dimer energy formula. This 
is similar to Fermi gases in non-trivial confinement \cite{valiente2012}.

\section{Thomas-Efimov effect with a squeezed dimension} 
The zero-range three-boson equation with a periodic dimension can 
be obtained as a generalization the Skorniakov and Ter-Martirosian (STM) equation \cite{skm}.
For completeness, let us first establish the STM equation for a regular 3D system
in homogeneous space (see Ref.~\cite{yamashita2013} for a discussion of the general 
form of the STM equations in 3D). 
\begin{small}
\begin{multline}
f(\vec q) =2\,\tau_{3D}\left(E_3-\frac34 q^2\right)
\int d^3p {f(\vec p)\over
E_3-q^2-p^2-\vec q\cdot \vec p},
\end{multline}
\end{small}
where $E_3$ is the three-body energy and $f(\vec q)$ is the so-called spectator
function which uniquely defines the three-body wave function \cite{yamashita2013}. 
As we only consider the equal mass case, there is only one spectator function 
(a detailed discussion can be found in Ref.~\cite{castin2011}). Here $\tau_{3D}$ 
is the two-body T-matrix given below in Eq.~\eqref{eq12}.

We may now split the momentum variable in the directions transverse to 
($p_\perp$ and $q_\perp$)
and parallel to ($p_z$ and $q_z$) the 'squeezed' dimension. This means that along the $z$-axis
the momenta have to be quantized as discussed above. The integral over all 
momenta in the STM equation thus becomes an integral over $p_\perp$ and a sum 
over the modes of the periodic dimension. Likewise, the kinetic energy 
(or free Green's function) is naturally split into continuous ($p_\perp$) and discrete
contributions ($p_z$). The generalized STM equation now becomes
\begin{small}
\begin{multline}
f(\vec q_\perp,n) =2\,\tau_R\left(E_3-\frac34(q_\perp^2+{n^2\over R^2})\right)
\\ \times \sum_m\int d^2p_\perp {f(\vec p_\perp,m)\over
E_3-q_\perp^2-p_\perp^2-\vec q_\perp\cdot \vec p_\perp-{n^2\over
R^2}-{m^2\over R^2}-{n\;m\over R^2}}, \label{zre1}
\end{multline}
\end{small}
where $\tau_R$ is the two-body T-matrix that we discuss momentarily
(see Eqs.~\eqref{rtau} and \eqref{rtau2}). 
The parameter that interpolates between two and three dimensions is $R$.
When $R$ approaches zero we are effectively in 2D and in the opposite limit
where $R\to \infty$ we go to the 3D case. 

We now discuss the Thomas collapse~\cite{th35} and Efimov
effect~\cite{ef70}. 
In momentum space the matrix elements of 
the Dirac $\delta$-function potential are constant. This introduces 
a singularity which must be resolved by proper regularization \cite{ad95} which
introduces a subtraction in the kernel 
of the zero-range three-boson integral equation, such that the properly 
regularized STM equations become
\begin{multline}
f(\vec q_\perp,n) = 2\,\tau_R\left(E_3-\frac34(q_\perp^2+{n^2\over R^2})\right) \\ \times 
\sum_{m}\int \frac{d^2p_\perp}{R} 
\left[g_{0R}(E)-g_{0R}(-\mu^2)\right] f(\vec p_\perp,m)
   \ \ , \label{zre5}
\end{multline}
where
\begin{small}
\begin{eqnarray}
g^{-1}_{0R}(E)=E-q_\perp^2-p_\perp^2-\vec q_\perp\cdot \vec p_\perp   
-\frac{n^2}{R^2}-\frac{m^2}{R^2}+\frac{n\;m}{R^2}.
\end{eqnarray}
\end{small}
Here $f(\vec p_\perp,m)$ is the momentum-space three-body
wave function that we need to determine.
The two-body T-matrix, $\tau_R$, is given by
\begin{align}
R \,\tau^{-1}_R(E)
 =\sum_n\int  
\frac{d^2p_\perp}{ E-
p_\perp^2-\frac{n^2}{R^2}}
-\sum_n\int  
\frac{d^2p_\perp}{E_2-
p_\perp^2-\frac{n^2}{R^2}}
\ , \label{rtau}
\end{align}
where $E<0$ and we chose the bound-state pole at $E_2$. Throughout most of this paper 
we will use units of $\hbar=\mu=M=1$, where $M$ is the boson mass and
$\mu$ is the momentum-space subtraction point of the regularization procedure \cite{ad95}. Performing the 
analytical integration over $\vec p_\perp$ and the sum, we have
 \begin{eqnarray}
\tau_R(E)=-R\left[
\pi\ln\left({\sinh\pi\sqrt{-E}R\over\sinh\pi\sqrt{-E_2}R}\right)\right]^{-1}
\ .\label{rtau2}
\end{eqnarray}
Taking into account that the T-matrix in 2D and 3D have different units, we 
recover the two limits via
\begin{eqnarray} \tau_{2D}(E)=\lim_{R\to 0} R^{-1}\tau_{R}(E)=-
\left[\pi\ln\left({\sqrt{-E}\over \sqrt{ |E_2| }}\right)\right]^{-1}, 
\label{eq9}
\end{eqnarray} 
which reproduces the standard 2D amplitude \cite{ad95}, and for 
$R\to \infty$ we obtain
\begin{eqnarray}
\tau_{3D}(E)=\lim_{R\to\infty}\tau_R(E) =\frac{1}{\pi^{2}}\left[\sqrt{E_2}-\sqrt{-E}\right]^{-1} 
\ , \label{eq12}
\end{eqnarray}
valid for $E<0$. For continuum energies,  $E>0$, the analytical extension in (\ref{eq9}) 
and (\ref{eq12}) can be performed from negative to the positive energy through the upper 
half of the complex energy plane.
There is a subtlety in Eq.~\eqref{zre5} since for any $R$  the kernel is noncompact if the 
subtraction term is disregarded.
Therefore, to take the 2D limit one has first
to regularize Eq.~\eqref{zre5} and then take the limit $R\to 0$. 
In addition, to get the famous 2D results of 
Bruch and Tjon \cite{tjon}, one has to also take the limit $E_2\to 0$. 

For $R\to 0$, the Efimov limit given by $E_2\to 0$ disappears because 
the homogeneous Eq.~\eqref{zre5} reduces to its usual 2D form. 
In this case, only 
states with $n=0$ are relevant. For higher $n$ the kinetic 
energy blows up and this makes terms with $n>0$ in the kernel of 
the bound state equation irrelevant. 
The 3D Thomas collapse (infinitely negative ground state three-body energy) occurs
when letting $\mu\to \infty$
in Eq.~\eqref{zre5} after taking $R\to \infty$.
For any finite $R$, the three-body ground
state still collapses for $\mu\to \infty$ as the kernel of Eq.~\eqref{zre5} 
is noncompact if the subtraction term is ignored. 
Here we will
work in the limit where $\mu$ is finite and we take units 
such that $\mu=1$ as discussed above. 
We note that the equivalence between
the Thomas and Efimov effects that happens in 3D \cite{adh88} is
broken with a compact dimension.

\section{Trimer energy with periodic boundary conditions} 
\begin{figure}[!thb]
\centering
\includegraphics[scale=0.30,clip=true]{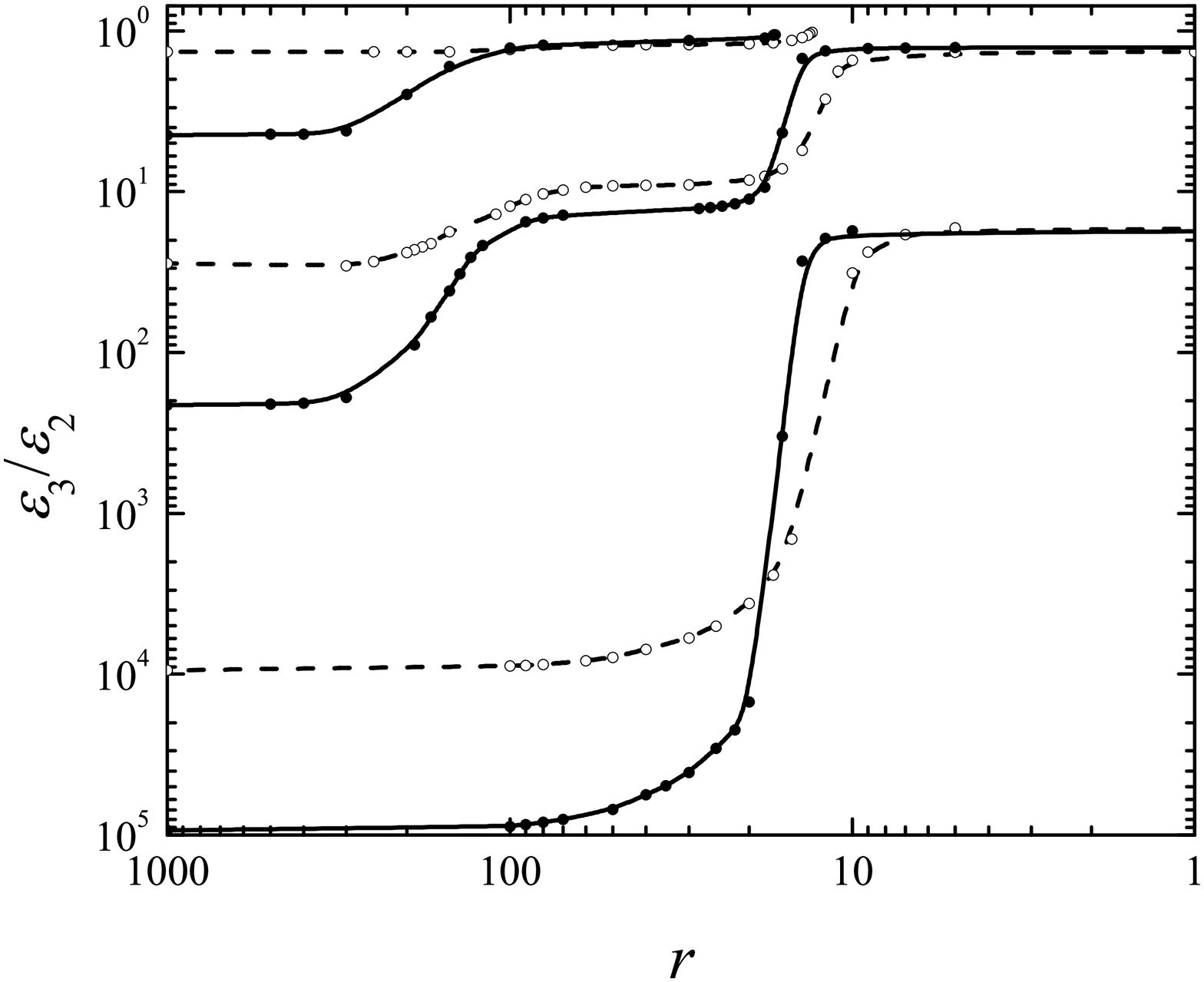}
\caption{$\epsilon_3/\epsilon_2$ as a function of $r$, for $\epsilon_2^{3D}=$ $10^{-7}$ 
(full circles) and $10^{-6}$ (empty circles). The solid and dashed lines are guides 
to the eye. As we approach the 2D limit ($r\to 0$), higher 
excited states disappear and only the ground and first excited states remain. Note that 
the values of $r$ and $\epsilon_3/\epsilon_2$ increase from right to left and top to 
bottom respectively.}
\label{compact}
\end{figure}

We now present the numerical solution of the 
trimer bound state equation in Eq.~\eqref{zre5}.  
The calculations presented are done for two different
values of $E_2$ which are independent of $R$ for 
reason of numerical convenience. 
In an experiment 
this could be achieved by using a magnetic Feshbach resonance
and changing the field along with the trap size in such a 
manner that $E_2$ remains fixed. 
We will address 
shortly what qualitative changes we expect for the
case where $E_2$ varies with $R$ as we have discussed 
in the dimer section above.
Introducing explicit units for clarity,
the dimensionless quantities we will use are $\epsilon_3=E_3/E_0$, $\epsilon_2=E_2/E_0$ 
and $r=R\,\mu/\hbar$, where $E_0=-\frac{\hbar^2\mu^2}M$.
In order to explore the dimensional crossover transition, Fig.~\ref{compact} shows the ratios 
$\epsilon_3/\epsilon_2$ as a function of $r$ for the ground, first, and second
excited states. Note that the last state goes into the continuum before the 
2D limit is reached.

The points are calculated for two fixed two-body 
energies $\epsilon_{2}^{3D}=10^{-6}$ (empty circles/dashed lines) 
and $10^{-7}$ (full circles/solid lines). In a pure 3D calculation
these parameters are on the $a>0$ side of the resonance and one finds three three-body bound states. 
The points at which we calculated the energies are 
shown explicitly, while the curves are guides to the eye.
For the largest $r=1000$, the energies were obtained from a pure 3D STM equation.
The plot shows a very interesting dimensional crossover result. We have one sharp transition 
for the ground state and two for the first excited state, while the second excited state
has two transitions before it disappears. This plot is reminiscient of 
the famous binding energy plot of Efimov trimers going from $a=\infty$ and into the 
three-body continuum \cite{ef70}. The sharp jumps may also be interpreted
as a sort of avoided crossing behaviour as $R$ changes. This is very similar
to what is seen in other confined systems such as short-range interacting 
two-body systems in a harmonic trap \cite{busch1998}. The latter show similar
jumps as function of the ratio of the interaction and confinement energy
scales and have been related to the Zeldovich rearrangement effect seen 
for particles in strong magnetic fields \cite{farrell2012,zinner2012}. 

The jumps can be understood by
considering the size of the trimer. It is given roughly by $\bar{r}\sim1/\sqrt{\epsilon_3}$. 
For $\epsilon_2=10^{-7}$ , from the ground state plateau at $\epsilon_3/\epsilon_2=93330$ for $r=1000$ and first 
excited state plateau at $\epsilon_3/\epsilon_2=211.79$ also for $r=1000$. From 
the energies in the 3D limit, we predict that for $\bar{r}=217.29$ the 
ground state trimer has a size that matches the size of the squeezed 
dimension, $r$. For the first excited state the corresponding number is
$\bar{r}=10.35$. Looking at Fig.~\ref{compact} we see that these 
numbers match quite well with the values calculated numerically. 
This allows us to interpret the jumps as signaling that the 2D
limit is reached first for the ground state and then for the 
first excited state. Due to the additional avoided crossing, 
we get the second jump for the first excited state after which 
it is forced to go to the only excited state that is present in 
the strict 2D limit for $r\to 0$.
The same 
analysis can be made for $\epsilon_2=10^{-6}$ with $\bar{r}=10.27$ and $\bar{r}=188.98$, 
respectively, for the ground and first excited state. Varying $r$ from large to small 
values (left to right), 
the 3D$\rightarrow$2D transition occurs for $r\sim10$, where we have the 
disappearance of the higher (second in our case) excited states in order to reproduce the well known 
2D results with two bound states proportional to $\epsilon_3/\epsilon_2=16.52$ 
and $\epsilon_3/\epsilon_2=1.27$ \cite{tjon}.

From the experimental point of view it may be difficult to keep the dimer energy, $E_2$, 
constant. However, the transitions observed in Fig.~\ref{compact} 
will not disappear due to a variation of $\epsilon_2$ with $r$. A change in the 
dimer energy does not move the jumps significantly. Larger dimer energies will cause  
the 3D plateau to move to lower $\epsilon_3/\epsilon_2$ ratio and push the beginning of the transition 
to smaller $r$, thus making the transition region broader. In cases where there are
four or more states in the spectrum, the higher states (second excited and above)
will go to the continuum before one reaches the 2D limit. Whether they show two
plateaus depends on whether they enter the spectrum above or below the 
values $\epsilon_3/\epsilon_2=16.52$ and $\epsilon_3/\epsilon_2=1.27$ in 
analogy to what we see in Fig.~\ref{compact}. In general, 
the jumps happen when 
a state is commensurate with the energy of the transverse squeezed
dimension, while the associated plateaus are given by the 2D limit.

As an outlook we may consider a mass imbalanced system where the 
2D limit can be much more rich with many bound states \cite{bellotti2D,ludo2010}. 
This immediately implies that there could be more plateaus for these
systems and that the sequence of jumps will be more involved but 
potentially even more interesting. We leave this issue for future
studies.

The authors thank partial support from the Brazilian agencies
FAPESP (2013/04093-3), CNPq and CAPES (88881.030363/2013-01), and by the Danish
Council for Independent Research DFF Natural Sciences
and the DFF Sapere Aude program. We thank 
M. Valiente and J. Levinsen for enlightening discussions

\end{document}